\documentclass{PoS}
\usepackage{float}
\usepackage[caption = false]{subfig}

\title{Intrinsic time lags in blazar flares and the search of Lorentz Invariance Violation signatures}

\ShortTitle{Intrinsic time lags in blazar flares and the search of LIV signatures}

\author{\speaker{C. Perennes}\\
       Sorbonne Universit\'es, UPMC Universit\'e Paris 06, Universit\'e Paris Diderot, Sorbonne Paris Cit\'e, CNRS, Laboratoire de Physique Nucl\'eaire et de Hautes Energies (LPNHE), 4 place Jussieu, F-75252, Paris Cedex 5, France\\
        E-mail: \email{cedric.perennes@lpnhe.in2p3.fr}}
        
\author{H. Sol\\
        LUTH, Observatoire de Paris, PSL Research University, CNRS, Universit\'e Paris Diderot, 5 Place Jules Janssen, 92190 Meudon, France\\
        E-mail: \email{helene.sol@obspm.fr}}

\author{J. Bolmont\\
         Sorbonne Universit\'es, UPMC Universit\'e Paris 06, Universit\'e Paris Diderot, Sorbonne Paris Cit\'e, CNRS, Laboratoire de Physique Nucl\'eaire et de Hautes Energies (LPNHE), 4 place Jussieu, F-75252, Paris Cedex 5, France\\
        E-mail: \email{bolmont@lpnhe.in2p3.fr}}

\abstract{Some Quantum Gravity models predict a violation of Lorentz invariance. Namely, the velocity of photons in vacuum could depend on their energies. One possibility for Lorentz Invariance Violation (LIV) searches is to look for energy-dependent delays in the arrival time of very high energy (VHE) photons coming from distant sources such as TeV emitting Active Galactic Nuclei (AGN), mainly blazars. Up to now, observations of flaring AGN have only provided upper limits on LIV energy scale since no significant time-delay was found and confirmed. 

However, AGN are not perfect sources for such LIV searches because intrinsic temporal effects can be produced by emission processes, similarly to what has been observed with gamma-ray bursts. With the beginning of the Cherenkov Telescope Array (CTA) operations in the coming decade, significant lags should be measured and the question of the origin of these lags will arise. In particular, propagation effects (LIV) will have to be disentangled from source-intrinsic effects. A precise time-dependent modeling of blazar flares becomes necessary to understand the different origins of these time-delays.

In this contribution we report on time-delay studies for blazars in the VHE domain. Using different time-dependent emission scenarios of blazar flares, we illustrate their resulting intrinsic spectral lags and provide tools for further analyses of the fine temporal and spectral behavior of such astrophysical sources. Tightly constraining intrinsic effects should then allow to better highlight any extrinsic contribution due to LIV. Conversely, any significant detection of a time-delay should provide new constraints on emission scenarios.}

\FullConference{35th International Cosmic Ray Conference --- ICRC2017\\
		10--20 July, 2017\\
		Bexco, Busan, Korea}

\begin{document}

\section{Introduction}
Several tentative approaches to Quantum Gravity indicate a possible violation of Lorentz Invariance through modified, non-trivial, dispersion relations for photons in vacuum (see \textit{e.g.} \cite{Amelino2013} for a review). This leads to the introduction of a model-independent test theory where the dispersion relation is expressed simply as
\begin{equation}
\label{eq:disprel1}
E^2 \simeq p^2 c^2\times\left[1 \pm \sum_{n=1}^\infty k_n\ \left(\frac{E}{E_P}\right)^n\right],
\end{equation}
where $c$ is the low energy limit of the speed of light, $E_P$ is the Planck scale ($1.22\times10^{19}$ GeV), and $k_n$ coefficients to be measured, or constrained. The sign $\pm$ in this equation takes into account the possibility to have subluminal or superluminal effects. The possibility to observe an effect on propagation of very high energy (VHE) photons from distant astrophysical sources, such as Gamma-ray Bursts or flaring Active Galactic Nuclei (AGN) was first proposed in the late 90s \cite{Amelino-Camelia1998}.

From the dispersion relation above, the delay between two photons emitted \textit{at the same time and location} by an astrophysical source at redshift $z$ and with energies $E_h > E_l$ can be expressed as
 \begin{equation}
\label{eq:timez5}
\Delta t_n \simeq \pm\,\frac{n+1}{2}\,\frac{E_h^n - E_l^n}{E_{QG}^n} \int_0^z \frac{(1+z')^n}{H(z')} dz'.
\end{equation}
where $H(z) = H_0 \sqrt{\Omega_m\,(1+z)^3 + \Omega_\Lambda}$ is the Hubble parameter and where $E_{QG}$ is a parameter to be constrained, expected to be related to some Quantum Gravity energy scale, presumably not too far from the Planck scale. Only the first order ($n = 1$) is within reach of present-day gamma-ray detectors while order $n = 2$ will be better constrained by the future observatory CTA (\textit{Cherenkov Telescope Array} \cite{Actis2011,Acharya2013}). Up to now, the parameter $\Delta t_n$ has always been found to be compatible with zero, except in one occasion dicussed below. The best limits obtained so far for $n = 1$ are above the Planck scale while for $n = 2$ they reach $\sim5\times10^{10}$ GeV (See Table 2.2, page 66 of \cite{bolmont:tel-01388037} for a summary of all results obtained until 2016).

The key assumption used to obtain Eq.~\ref{eq:timez5} is that high and low-energy photons are emitted at the same time and place and are not affected by other propagation effects during the radiation transfer. This hypothesis neglects any delay that could be induced at the source by high-energy photon production mechanisms. For instance, energy stratification induced by particle acceleration at shock fronts and complex geometry of the VHE emitting zone can result in frequency-dependent  photons emission time \cite{Sokolov2004}, as well as the microphysics of the VHE radiation production itself as shown hereafter. Such effects need to be better described in order to distentangle them from any Lorentz Invariance Violation (LIV) delay.
Up to now, only one flare of Mkn~501, recorded on July 9, 2005 by MAGIC, was found to exhibit a lag of $4\pm1$ min between energy bands $<$250~GeV and $>$1.2~TeV \cite{Albert2007}. Despite rather low photon statistics, this example of a significant lag shows that time delays can exist in AGN flares and that a mixture of both LIV and intrinsic delay is also possible. 

The work presented here is a first step to describe and interpret intrinsic time delays through the use of a time-dependent model of AGN, focusing on gamma-ray emission. Section 2 introduces the model and its different components responsible for the time evolution. Section 3 presents the intrinsic time delays obtained from the model and their interpretations with two different scenarios. The last section discusses the implications of the model on LIV searches and give some prospects.

\section{Model description}  

The time dependent model used for this study is adapted from \cite{kat03}. The time evolution is coming from the variations of the particule distribution $N_e$, through the following kinetic equation :

\begin{equation}	
\label{equadiff}
\frac{\partial N_e(t,\gamma)}{\partial t} = \frac{\partial}{\partial \gamma} \left\{ \left[\gamma^2 C^{cool}(t) +
\left(C^{adiab}(t) - C^{acc}(t) \right)\ \gamma \right]\ N_e(t,\gamma) \right\}
\end{equation} 

where $C^{cool}(t)$ corresponds to radiative losses, through synchtrotron and Inverse-Compton (IC) emission, $C^{acc}(t)$ describes the acceleration proccesses and $C^{adiab}(t)$ characterises the adiabatic losses. The analytic solution found in \cite{kat03} is used, with the following coefficients : 
\begin{equation}
C^{cool}(t) = \frac{4\sigma_TU_B(t) \left(1 + \frac{1}{\eta}\right)}{3m_ec} \hspace{0.75cm} ; \hspace{0.75cm} C^{acc}(t) = A_0 \left( \frac{t_0}{t} \right)^a \hspace{0.75cm}; \hspace{0.75cm} C^{adiab}(t) = \frac{1}{t}
\end{equation}	
where $U_B(t) = B(t)^2/(8\pi)$ corresponds to the magnetic field energy density, $\eta$ is the ratio between $U_B(t)$ and the synchrotron radiation density and represents the IC energy losses as a fraction of the synchrotron energy losses. This hypothesis is needed in order to find an analytic solution to equation (\ref{equadiff}). $A_0$ is the initial acceleration coefficient and $t_0$ is the initial time, also used as the time scale of the evolution for all parameters, defined as $t_0 = R_0/V_{exp}$ with $R_0$ the initial radius of the emission zone and $V_{exp}$ the expansion velocity corresponding to the adiabatic losses process. This parameter $t_0$ defines the evolution time  of the following time-dependent parameters : 
\begin{equation}
B(t) = B_0 \left( \frac{t_0}{t} \right)^b \hspace{1cm} ;\hspace{1cm}  R(t) = V_{exp} \times t = R_0\left( \frac{t_0}{t} \right)
\end{equation}
where $B_0$ is the initial magnetic field strengh. The initial condition of the particle distribution is described by a power law function with a cut-off : 
\begin{equation}
N_e(t_0,\gamma) = K_0 \gamma^{-n} \left[ 1 - \left( \frac{\gamma}{\gamma_{cut}}\right)^{n+2} \right]
\end{equation}
with $K_0$ the density of electron at $\gamma = 1$ in unit of cm$^{-3}$, $n$ is the slope of the power law function and $\gamma_{cut}$ is the energy where the high-energy cut-off occurs.

The solution given above allows to get a set of electron spectra, corresponding to the evolution of the initial injected electron spectrum. Then to obtain the corresponding evolution of the Spectral Energy Distribution (SED), a one-zone homogeneous Synchrotron Self-Compton (SSC) model is used for each electron spectrum obtained thanks to the analytic solution (see \cite{kat01} for more details on the model). 

\section{Results and interpretations}

\begin{table}
	\begin{center}
		\begin{tabular}{|c| c c c | c| c c c |} \hline
			$z$ 				& 0.03 			& 				& 			&$B_0$	& 0.64 / 0.9 	&					& G			\\
			$\delta$ 		& 40 			& 				& 			&	$b$ & 1 		& 					& 			\\
				 			& 	 			&				& 		 	&$\eta$ & 3 		& 					& 			\\
			$K_0$ 			& 200 			&				& cm$^{-3}$ 	&$A_0$ 	& 4.5 	& $\times 10^{-4}$	& s$^{-1}$	\\
			$\gamma_{cut}$ 	& 1 				&$\times 10^{6}$& 			&$a$ 	& 4.5	& 					&	 		\\
			$n$				& 2.4			& 				& 			&$R_0$ 	& 9.4 	& $\times 10^{14}$ 	& cm			\\
			\hline
		\end{tabular}
	\end{center}
	\caption{List of parameters used for scenario 1 and 2. The two scenarios share a common set of parameters except for $B_0$ where both values are mentionned.}
	\label{param}
\end{table}

\subsection{Methodology}
In order to study the basic time-evolution, the model is simplified by neglecting adiabatic losses. Only acceleration and radiative cooling processes are kept and will contribute to the time evolution, either by modification of the electron spectrum or evolution of the magnetic field with time.
The remaining parameters are $z$, $\delta$, $K_0$, $\gamma_{cut}$, $n$, , $B_0$, $b$, $\eta$, $A_0$, $a$, $R_0$. The parameter $V_{exp} = c/\sqrt{3}$ is also kept for the definition of $t_0$ and its value is taken as the sound velocity in a relativistic plasma.

The model gives the evolution of the electron spectrum and the SED for each time-step. From the model, lightcurves can be extracted in different energy bands by integrating the flux over the energy. Then a method of cross-correlation \cite{Ede88} is used to measure a time-delay by comparing the lowest energy lightcurve to all the others. This allows to obtain the evolution of the time delay with respect to the energy. The lightcurves are taken in the energy range between 1 MeV, chosen in order to avoid the influence from synchrotron emission, and 50 TeV. This energy range is then splitted into 30 energy bands with equal width in logarithmic scale. Two scenarios were investigated using a common set of parameters given in Table \ref{param}, which are typical values for blazars.

\begin{figure}
		\includegraphics[width=0.5\linewidth]{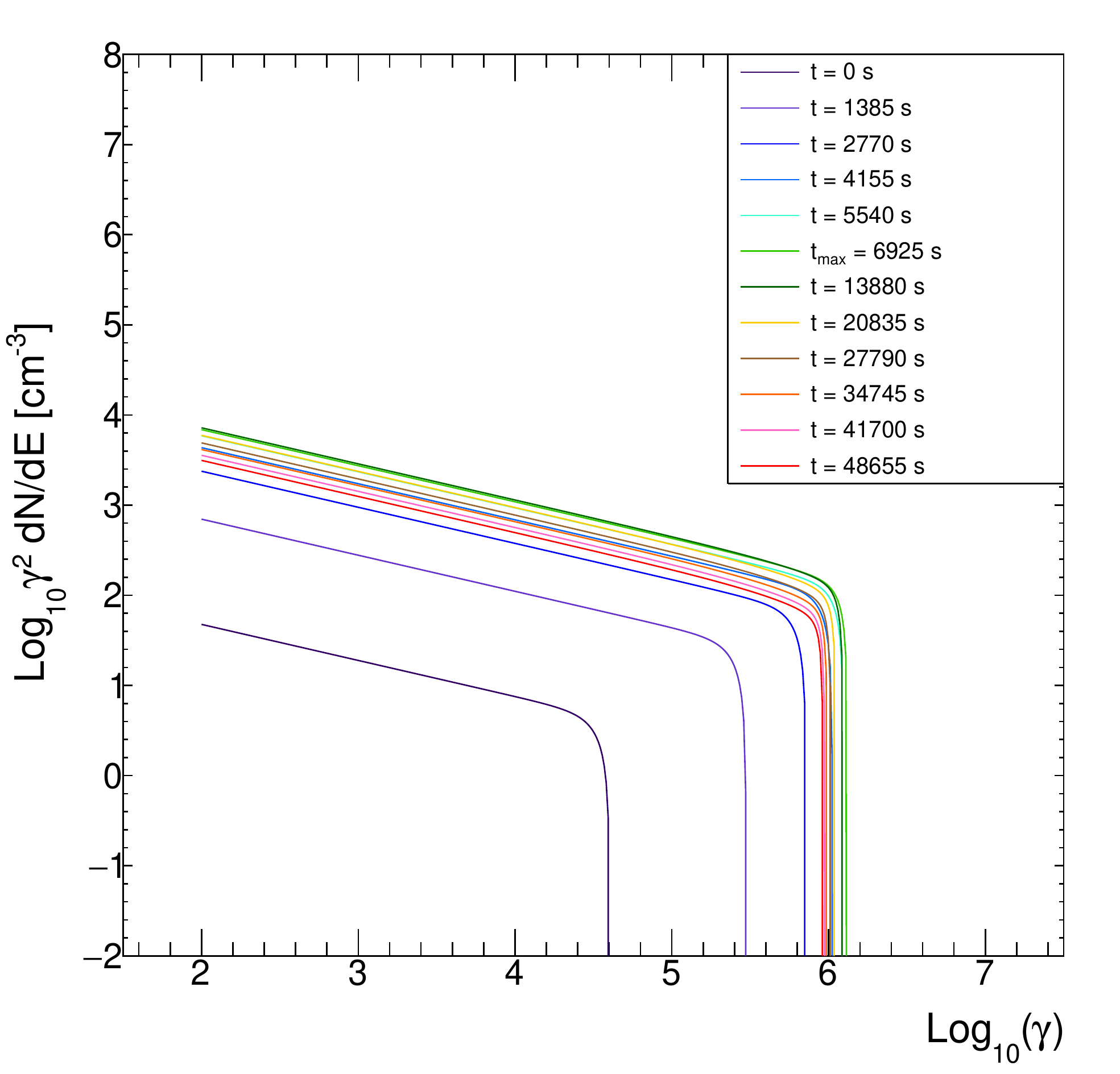}
		\hfill
		\includegraphics[width=0.5\linewidth]{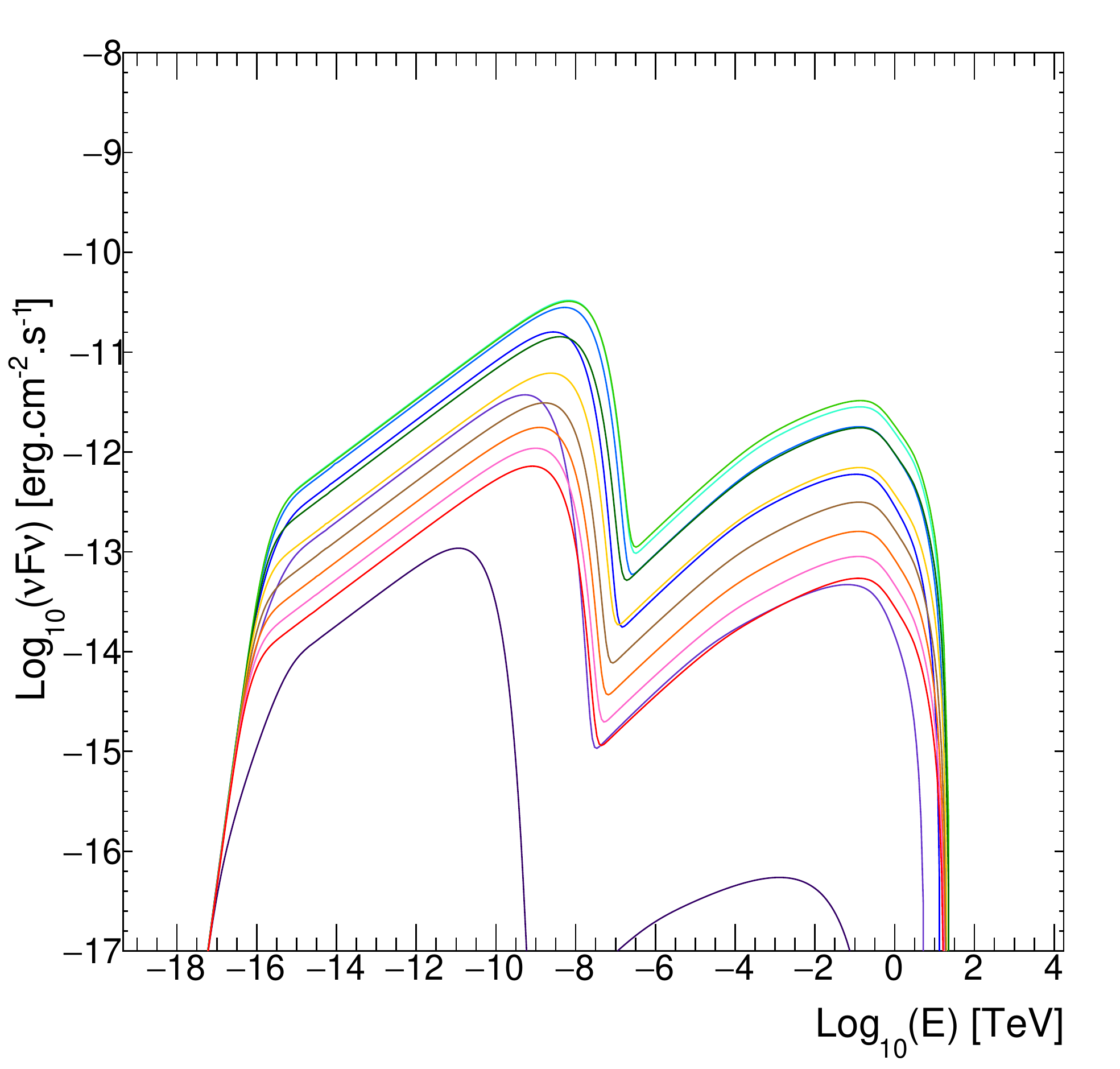}
		\caption{Results from the model for scenario 1. (Left) Electron spectrum evolution with time. (Right) Spectral energy distribution evolution including absorption by standard extragalactic background light. The time corresponds to the observed time on Earth taking into account Doppler boosting. The two plots share the same color code. The time $t_{max}$ corresponds to the time when the highest value of $\gamma_{max}$ for the electron spectrum is reached.}
		\label{sed:1}
\end{figure}

\begin{figure}
\subfloat[Scenario 1]{
\includegraphics[width=0.5\linewidth]{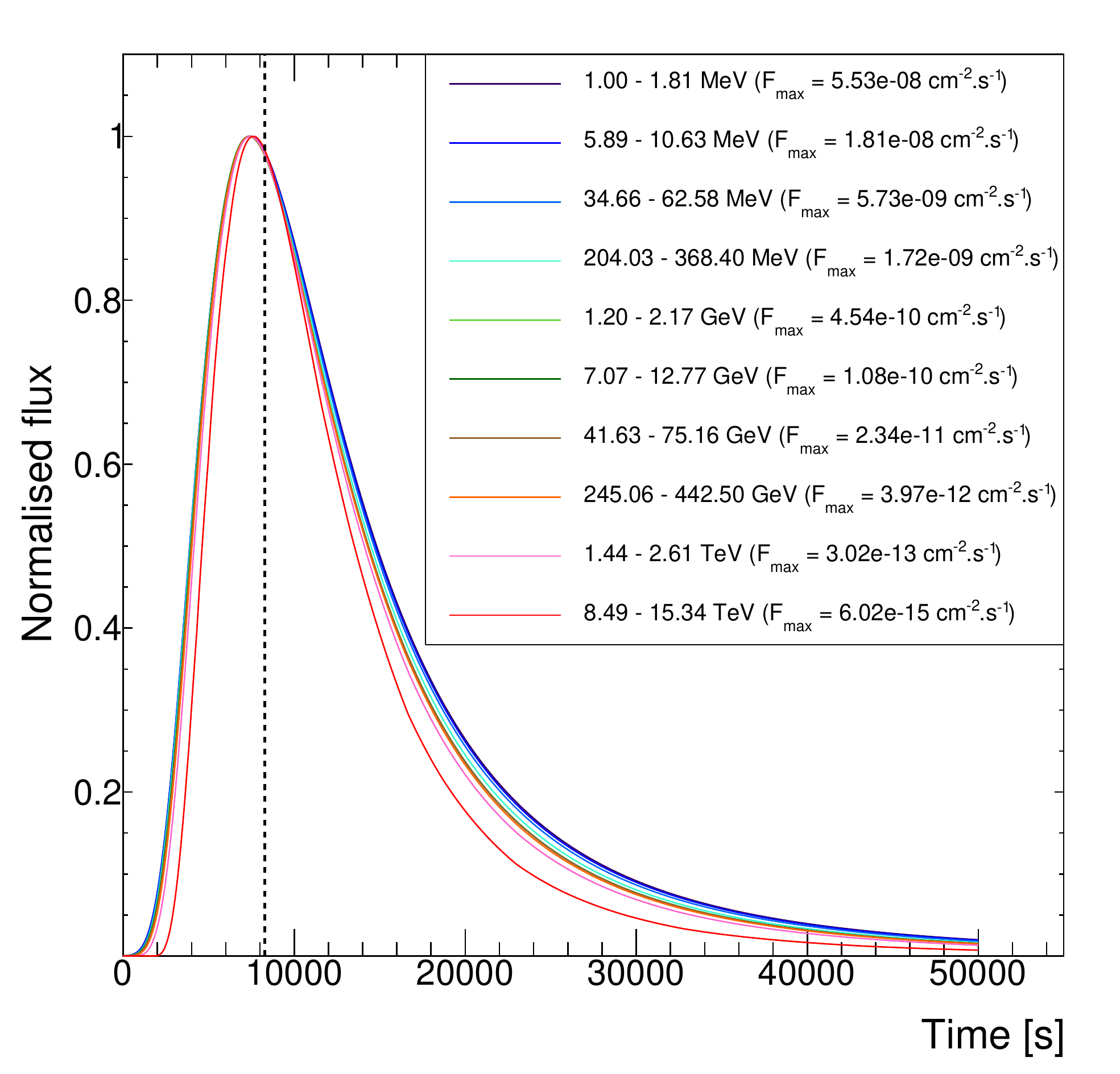} 
\hfill
\includegraphics[width=0.5\linewidth]{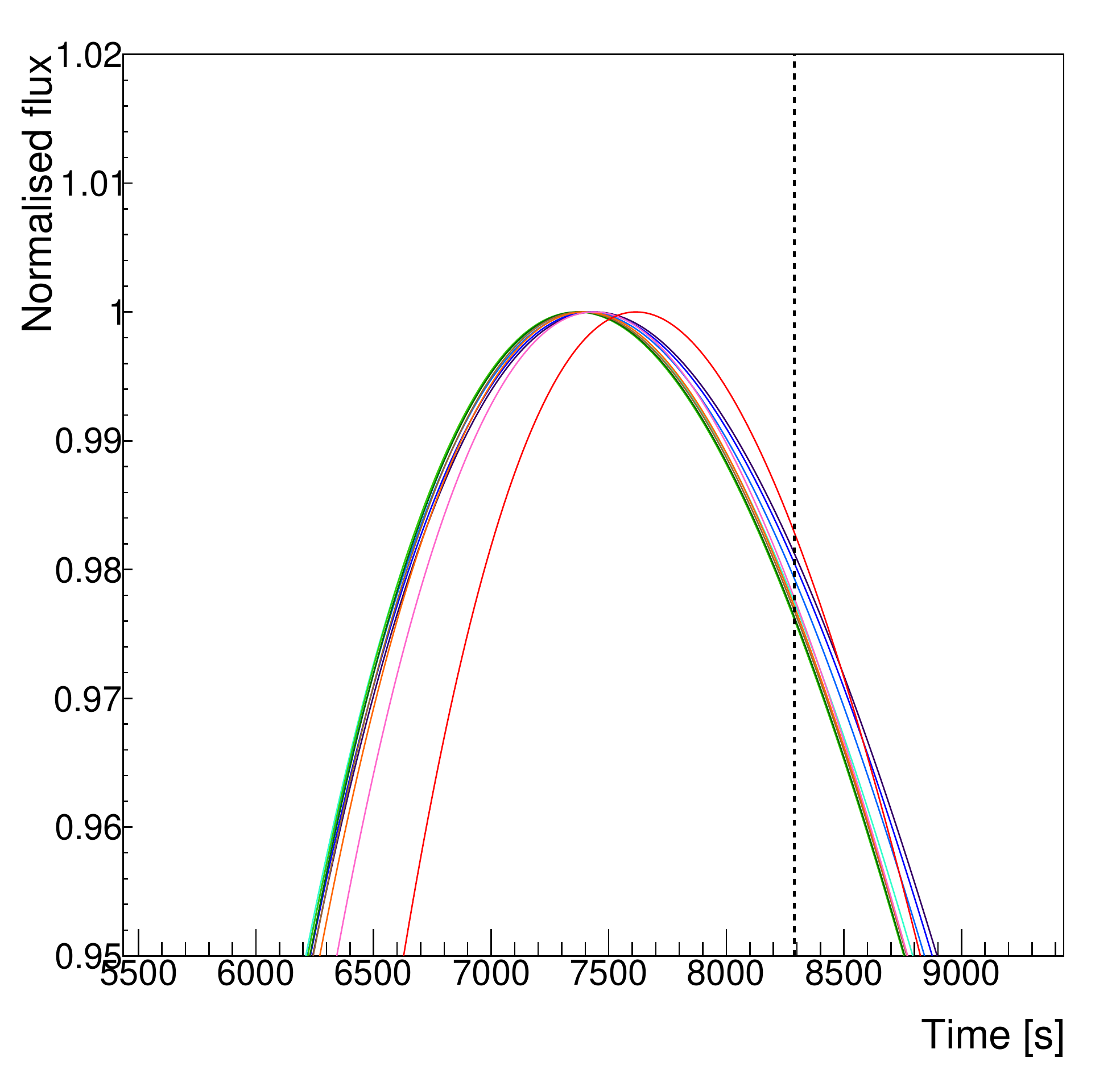}
		\label{lc:1}
}
\\
\subfloat[Scenario 2]{
		\includegraphics[width=0.5\linewidth]{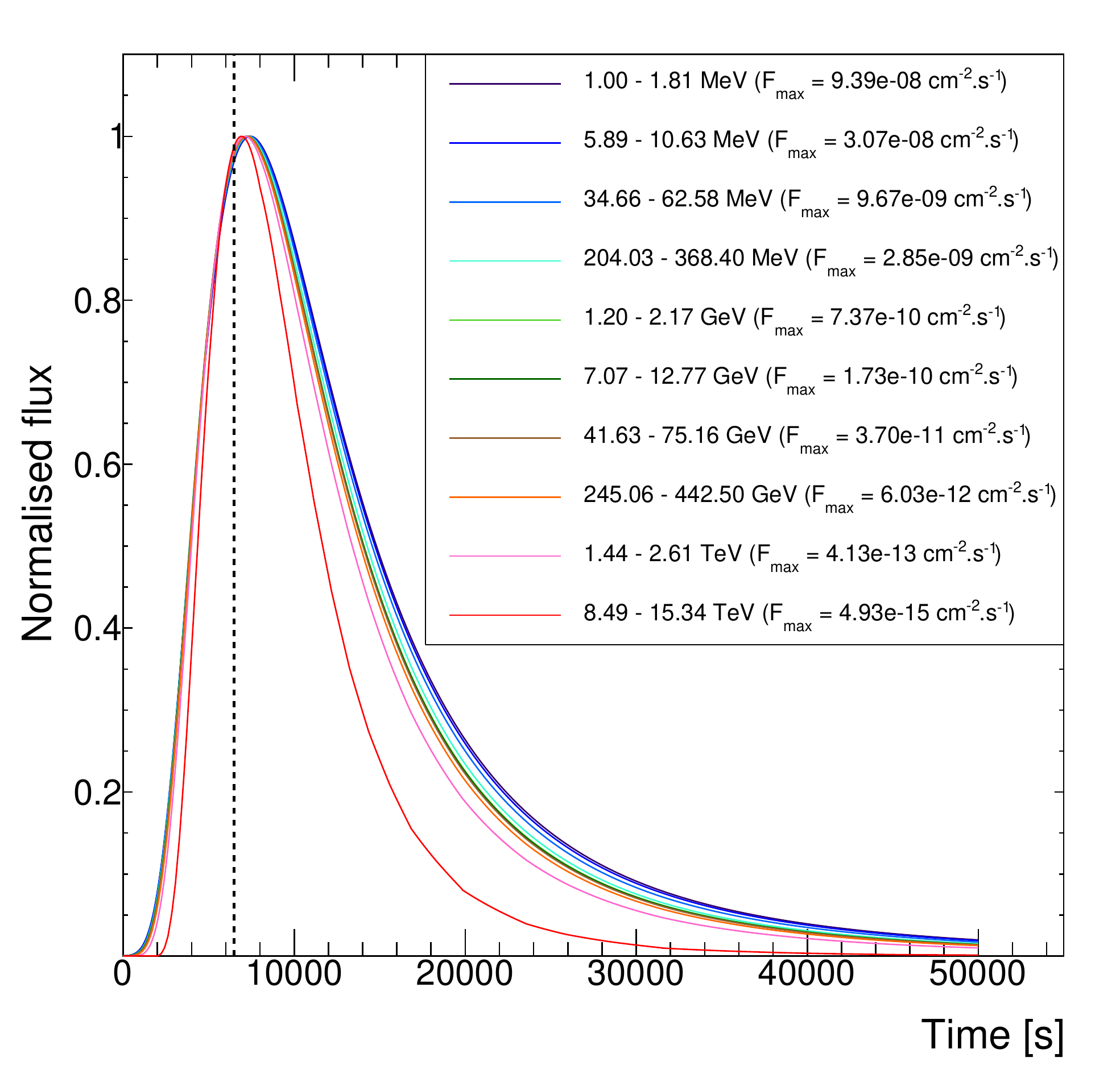}
		\hfill
		\includegraphics[width=0.5\linewidth]{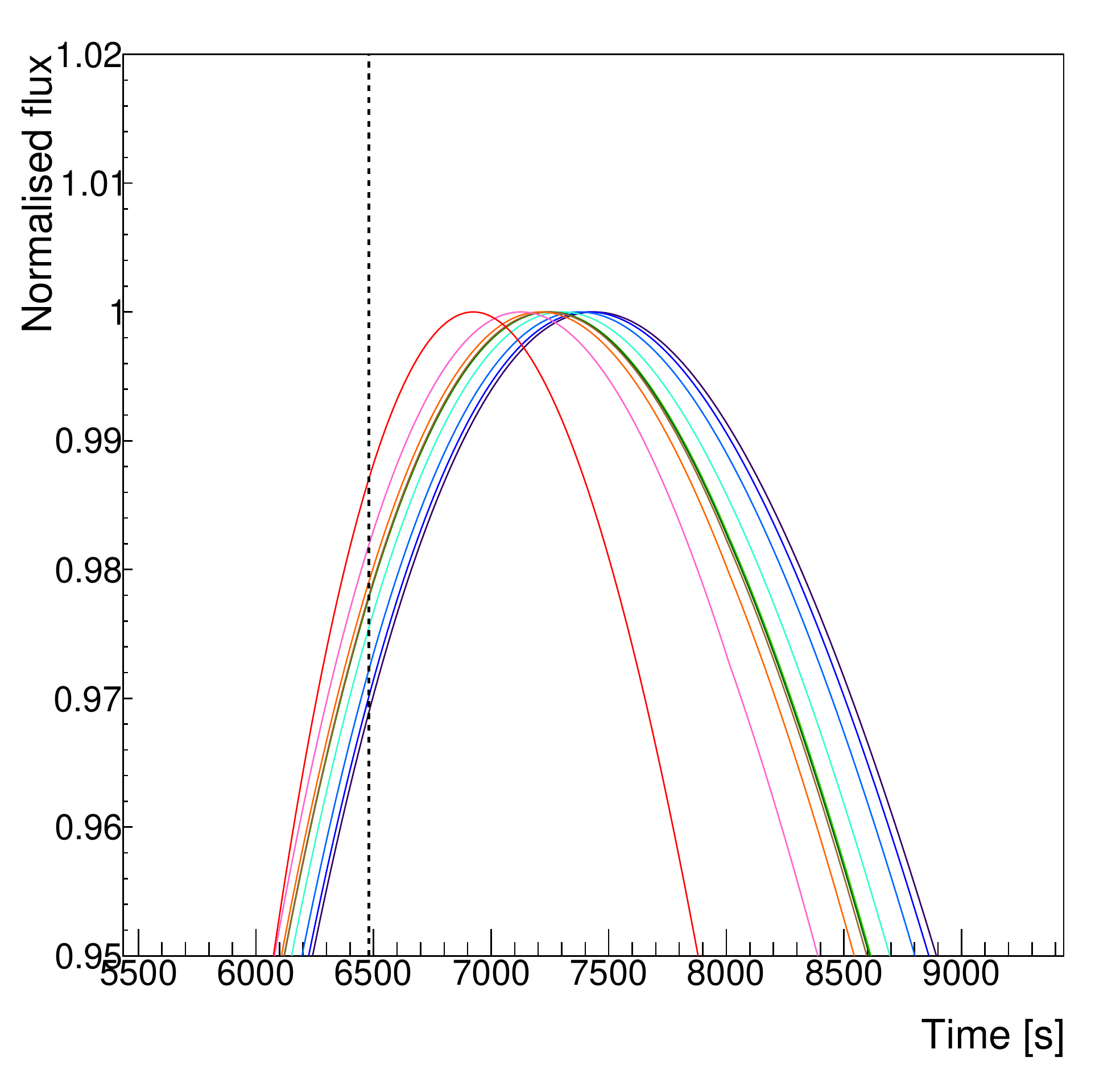}
		\label{lc:2}
}
		\caption{Normalised lightcurves obtained from the SED in different energy bands for both scenarios. (Left) Full time range. (Right) Zoom around the maximum of the flare showing a time-delay between the different energy bands. The vertical dashed line corresponds to the time when electrons reach their maximum energy $\gamma_{max}$ and show the moment when radiative cooling dominate over the acceleration at the energy $\gamma_{max}$.}
\end{figure}

\subsection{Scenario 1 : Long lasting acceleration scenario}
The resulting electron spectrum and SED evolution for scenario 1 are shown in Figure \ref{sed:1}. The evolution of the electron spectrum shows a fast acceleration phase with an increase of the electron density and of $\gamma_{max}$  followed by a slow cooling phase. The correspoding SEDs show similar phases leading to a a flare. The time when the maximum value of $\gamma_{max}$ is reached during the evolution is defined as $t = t_{max}$ and is reported on the lightcurves in Figure \ref{lc:1}. One can notice that all the lightcurves fall down before the time $t_{max}$, meaning that the electrons are still accelerated when the flux begin to decrease. The only parameter evolving with time being able to explain the fall down of the flux before the end of the acceleration phase is the evolution of $B(t)$ which decreases with time, reducing the overall SSC emission.

\begin{figure}
\includegraphics[width=0.5\linewidth]{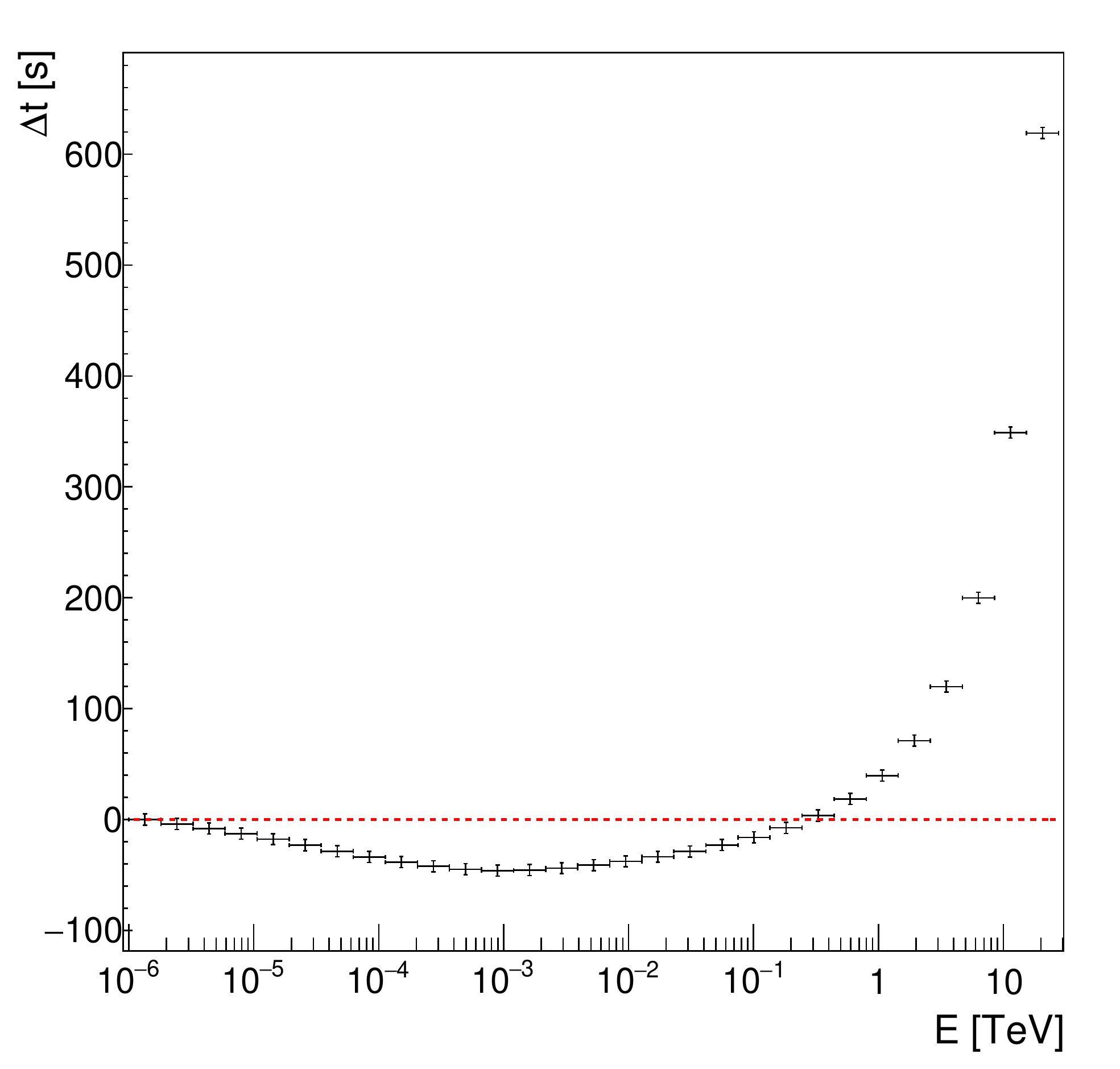}
\hfill
\includegraphics[width=0.5\linewidth]{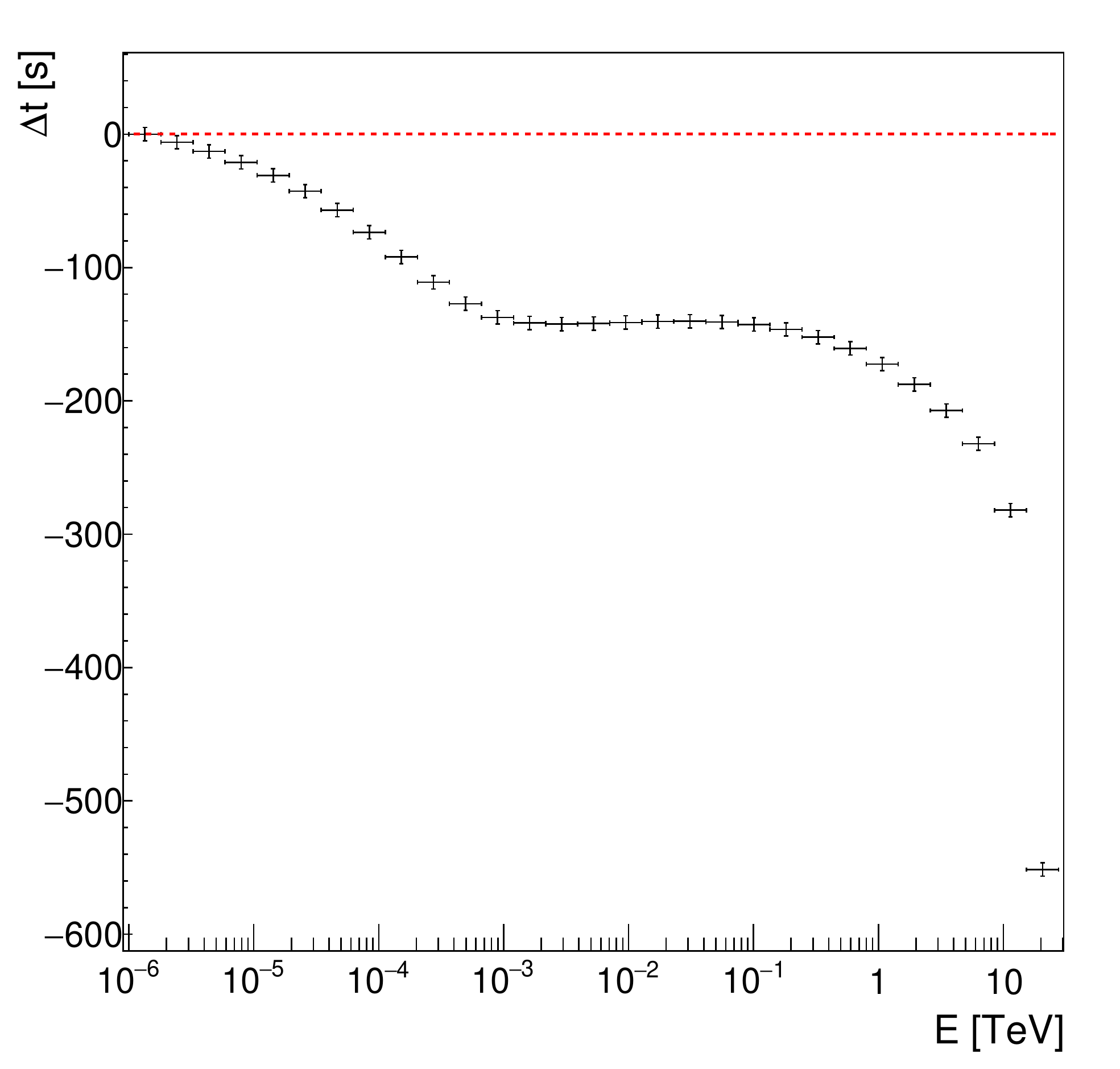}	
		\caption{Time delay between the lightcurve at $E_{ref} = 10^{-6}$~TeV with respect to the lightcurve at energy E for scenario 1 (left) and scenario 2 (right). The evolution of time delay shows two regimes separated at about the same energy in the two scenarios, $E \approx 10^{-3}$~TeV. The horizontal red dashed line correspond to no time-delay with respect to the lightcurve of energy $E_{ref}$.}
		\label{dt}
\end{figure}

A clear time delay is obtained as shown in Figure \ref{dt} (left) where two regimes can be distinguished. The first regime shows a positive time delay for high energies ($E > 10^{-3}$~TeV) which is easily explained by the time needed for the electrons to reach an energy capable to induce very high energy photons, resulting in the highest energy photons arriving later. The second regime for low energy ($E < 10^{-3}$~TeV), more difficult to explain, can be interpreted as due to the evolution of $B(t)$ dominating over the acceleration process. Indeed the lightcurves around the GeV range reach their maximum before the ones in the MeV range because the decrease of $B(t)$. The impact of this decrease affects less the edge of the IC bump and so the flare falls down ealier at intermediate energies

\subsection{Scenario 2 : Fast acceleration scenario}
A second scenario was investigated where the falling down of lightcurves happens after the time $t_{max}$. Compared to the previous one, this scenario a relatively better cooling efficiency in order to shorten the acceleration phase, involving a higher value for $B_0$. The lightcurves on Figure \ref{lc:2} show that the highest energies fall down first. One can notice that the flux is still increasing after the time $t_{max}$ involving that the cooling prevents electrons from reaching higher $\gamma_{max}$ but the electron density can still increase since the acceleration is still dominating over cooling for lower energy electrons.

The resulting time delay on Figure \ref{dt} (right) behaves differently from scenario 1 with also two regimes separated at the same energy $E \approx 10^{-3}$~TeV. The low energy regime is explained as scenario 1 by the evolution of $B(t)$ breaking the flux of the flare before the cooling of the electrons. The high energy regime is interpreted as an effect of the cooling time of the electrons. More energetic electrons have a shorter cooling time than low-energies ones leading to an ealier falling down for the high-energy lightcurves.

\section{Discussion and prospects}

The two scenarios presented here show two different behaviors of the time delay with respect to the energy. Using such information can bring new perspectives on AGN modeling. Depending on the scenario, a constraint on acceleration processes with respect to cooling processes  can be deduced since the difference between the two scenarios depend on when the cooling starts to dominate over acceleration of the electrons. Furthermore, measuring the value of time delays in different energy bands can also be used to model the time evolution of a source.

\begin{figure}
	\includegraphics[width=0.5\linewidth]{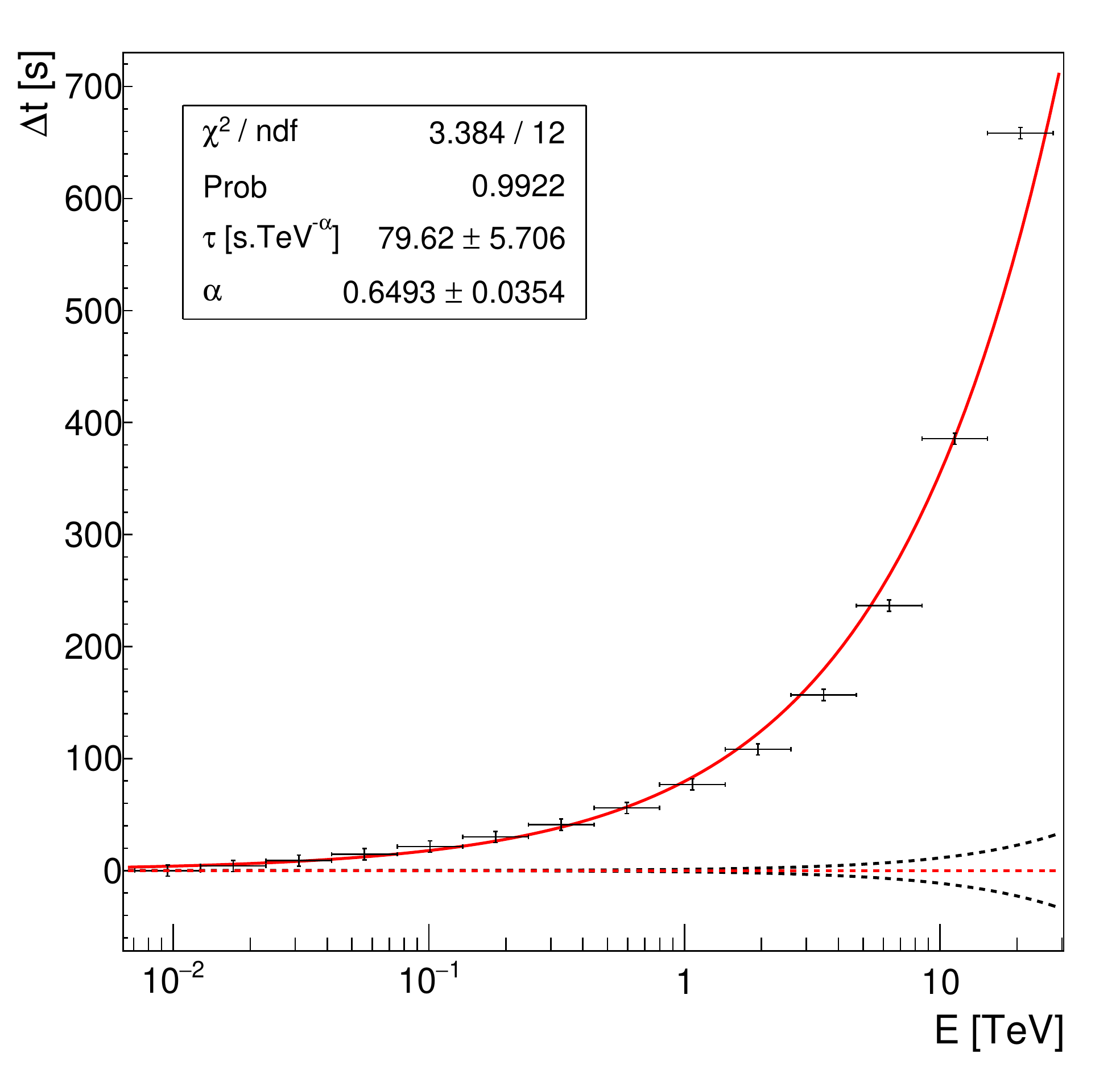}
	\hfill
	\includegraphics[width=0.5\linewidth]{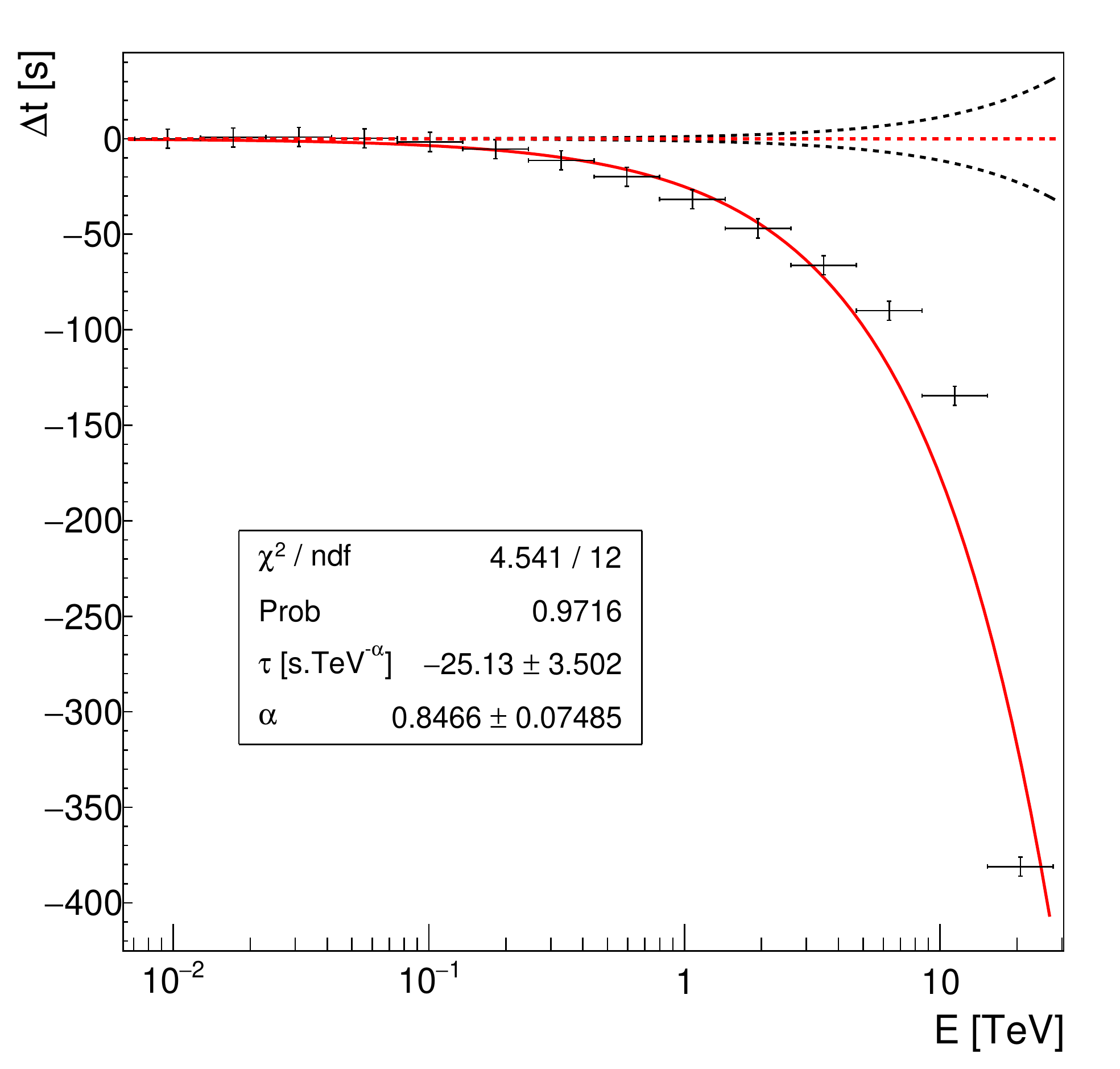}
	\caption{Intrinsic time delay between lightcurve at $E_{ref} \approx$ 30~GeV with respect to the lightcurve at energy E for scenario 1 (left) and scenario 2 (right). The red lines correspond to the result of a fit with a powerlaw function: $\Delta t = \tau \times E^{\alpha}$. The black dashed lines correspond to a time delay induced by LIV effects at the Planck scale for both subluminal ($\Delta$t > 0) and superluminal effects ($\Delta$t < 0), for z = 0.03.}
\label{dtgev}
\end{figure}

Investigations on LIV are currently done comparing the arrival time of low energy (GeV) and high energy (TeV) photons. In order to compare a LIV delay to the intrinsic one obtained with the proposed model, the cross-correlation method is used with a reference lightcurve at energies between 23.05 and 41.63 GeV. According to the model independent formalism of LIV Eq~.\ref{eq:disprel1}, the relation between LIV time delay and energy is expected to be a power law with an exponent 1 or 2. Figure \ref{dtgev} shows a similar behavior for intrinsic time delay, so the points were fitted with a power law function. The index are found to be different from the LIV effect value for both scenarios. This suggests to indicate a possible way to disentangle between intrinsic and LIV time delays. One can also notice that for these two scenarios, the absolute value of the intrinsic time delay is larger than the one expected from a LIV effect at the Planck scale. This is partially explained by the redshift chosen for the modeling (z~=~0.03), leading to a small value for propagation time delay according to Equation \ref{eq:timez5}, and also by the specific set of parameters chosen leading to the reported values of time delay. A more detailed analysis of these results is underway.

Up to now, only one observation of a significant time delay was reported at TeV energies \cite{Albert2007}. With CTA starting operations in the coming years, it is expected that more AGN flares will be detected with improved statistics and time resolution. It is then likely that significant lags will be measured. The interpretation of intrinsic lags and the way to distinguish them from LIV effects will be necessary. An approach such as the one proposed in this contribution could be used to interpret the lags as a combination of LIV and intrinsic effects. An essential difference between LIV and intrinsic effects is also that LIV effects should depend on the distance of sources (Eq.~\ref{eq:timez5}). This dependence will have to be exploited and population studies will be essential both for LIV searches and for the modeling side.


\begin{thebibliography}{99}
\bibitem{Amelino2013}
G. Amelino-Camelia,
\emph{Quantum-Spacetime Phenomenology},
\emph{Living Reviews in Relativity} {\bf 5}, 16, 2013

\bibitem{Amelino-Camelia1998}
 G. Amelino-Camelia, J. Ellis, N.~E. Mavromatos, D.~V. Nanopoulos and S. Sarkar,
\emph{Tests of quantum gravity from observations of gamma-ray bursts},
\emph{Nature} {\bf 393}, 793--765, 1998

\bibitem{Actis2011}
M. Actis, G. Agnetta, F. Aharonian and others,
\emph{Design concepts for the Cherenkov Telescope Array CTA: an advanced facility for ground-based high-energy gamma-ray astronomy},
\emph{Experimental Astronomy} {\bf 32}, 193-316, 2011
[{\tt arXiv:1008.3703}]

\bibitem{Acharya2013}
B.~S. Acharya, et al,
\emph{Introducing the CTA concept}, 
\emph{Astroparticle Physics} {\bf 43}, 3-18, 2013


\bibitem{bolmont:tel-01388037}
J. Bolmont,
\emph{Is the speed of light in vacuum really constant?},
\emph{Habilitation {\`a} diriger des recherches}, 2016,
[{\tt http://hal.upmc.fr/tel-01388037}]

\bibitem{Sokolov2004}
A. Sokolov, A.~P. Marscher, I.~M. McHardy,
\emph{Synchrotron Self-Compton Model for Rapid Nonthermal Flares in Blazars with Frequency-dependent Time Lags} 
\emph{ApJ} {\bf 613}, 725-746, 2004
[{\tt astro-ph/0406235v1}]

\bibitem{Albert2007}
J. Albert, E. Aliu, H. Anderhub and others,
\emph{Variable Very High Energy Gamma-Ray Emission from Markarian 501},
\emph{ApJ} {\bf 669}, 862-883, 2007
[{\tt astro-ph/0702008}]

\bibitem{kat03}
K. Katarzy\'nski, H. Sol \& A. Kus,
\emph{The multifrequency variability of Mrk 421},
\emph{A\&A} {\bf 410}, 101-115, 2003

\bibitem{kat01}
K. Katarzy\'nski, H. Sol \& A. Kus,
\emph{The multifrequency emission of Mrk 501},
\emph{A\&A} {\bf 397}, 809-825, 2001

\bibitem{Ede88}
R.~A. Edelson and J.~H. Krolik,
\emph{The discrete correlation function - A new method for analyzing unevenly sampled variability data},
\emph{ApJ} {\bf 333}, 646-659, 1988

\end{thebibliography}
\end{document}